# A mechanical autonomous stochastic heat engine

Marc Serra-Garcia*[1], André Foehr[1], Miguel Moleron[1], Joseph Lydon[1], Christopher Chong[2] and Chiara Daraio[1,3]

[1]Department of Mechanical and Process Engineering, Swiss Federal Institute of Technology (ETH), Zürich, Switzerland

[2]Department of Mathematics, Bowdoin College, Brunswick, ME 04011, USA

[3]Engineering and Applied Science, California Institute of Technology, Pasadena, California 91125, USA

*email: sermarc@ethz.ch

**Stochastic heat engines are devices that generate work from random thermal motion using a small number of highly fluctuating degrees of freedom. Proposals for such devices have existed for more than a century and include the Maxwell demon and the Feynman ratchet. Only recently have they been demonstrated experimentally, using e.g., thermal cycles implemented in optical traps. However, the recent demonstrations of stochastic heat engines are nonautonomous, since they require an external control system that prescribes a heating and cooling cycle, and consume more energy than they produce. This Report presents a heat engine consisting of three coupled mechanical resonators (two ribbons and a cantilever) subject to a stochastic drive. The engine uses geometric nonlinearities in the resonating ribbons to autonomously convert a random excitation into a low-entropy, nonpassive oscillation of the cantilever. The engine presents the anomalous heat transport property of negative thermal conductivity, consisting in the ability to passively transfer energy from a cold reservoir to a hot reservoir.**

Thermodynamics in low-dimensional systems far from equilibrium is not well understood, to the point that essential quantities such as work[1] or entropy[2] do not have a universally valid definitions in such systems. Formulating a physical theory for thermal processes in low dimensional systems is the subject of stochastic thermodynamics[3], an emergent field that has resulted in the discovery of a variety of microscopic heat engines[4-7], fluctuation theorems[8-11] and provided new insights on the connection between information and energy[5,12-14]. A central problem in stochastic thermodynamics is the construction and analysis of stochastic heat engines, the low dimensional analogs of conventional thermal machines. A stochastic heat engine is a device that operates between two thermal baths at different temperatures, and is able to produce work while suppressing the randomness inherent in thermal motion[1,4-7,15,16] [17]. Thermal engine operation is characterized by the presence of nonpassive states of motion, which have lower entropy (for the same energy) than equilibrium states[18,19] and therefore allow the extraction of energy without an associated entropy flow[1]. The interest in stochastic heat machines is motivated by the desire to understand energy conversion processes and limits at the fundamental level. This understanding, coupled with modern nanofabrication techniques is expected to result in more efficient and powerful thermal machines.

The concept of the stochastic heat engine dates back to the classical thought experiments of the Maxwell demon[13,20] and the Feynman ratchet[21,22]. Only very recently have working experimental realizations of the stochastic heat engine been reported on[4-7]. The bulk of these experimental realizations is based on the manipulation of a particle in an optical trap, and include the implementation of various adiabatic processes[7], active feedback mechanisms[5], as well as the realization of Stirling[6] and Carnot[4] cycles. These engines are nonautonomous since they operate under externally prescribed cycles. As a consequence, the energy they require to operate is orders of magnitude higher than the work they produce, and the externally prescribed dynamical cycle masks the significant challenges that hinder the description of autonomous physical systems[1]. In this Report, we describe a classical mechanical system that realizes autonomous thermal engine operation. Our proposed engine consists of two coupled ribbons and a cantilever beam connected to one of the ribbons (see Supplementary

Information for additional details). The nonlinearity in the ribbons dynamically and autonomously adjusts the coupling to the hot and cold thermal baths, and removes the need for the external control system used in non-autonomous engines. We demonstrate this concept on a table-top setup, which attains a non-passive state when driven by external noise. The Supplementary Information provides a discussion on the scaling of the system to microscopic dimensions, where Brownian motion is sufficient to induce measurable amounts of vibration.

The thermal cycle of our engine is analogous to the classical Stirling cycle (Figs. 1A and 1B), which consists of four steps performed on a working fluid -- heating, expansion, cooling and compression. The ribbon attached to the cantilever (which we label the main ribbon, and denote its displacement by $x_M$) plays the role of the working fluid. This ribbon is in contact with a cold thermal bath at temperature $T_C$. The cantilever acts as a piston that introduces cyclic compressions and expansions of the ribbon and extracts work from the fluctuations in the ribbon's tension. Due to geometric nonlinearity, this tension increases proportionally to the ribbon's vibrational energy and is analogous to the pressure of the gas in a conventional engine. A hot thermal bath, at temperature $T_H$ introduces the thermomechanical noise that causes Brownian motion. This heat bath is applied to a secondary ribbon (labeled $x_H$). The secondary ribbon is weakly coupled to the main ribbon and regulates the coupling between the hot reservoir and the main ribbon (Figs. 1C and 1D)

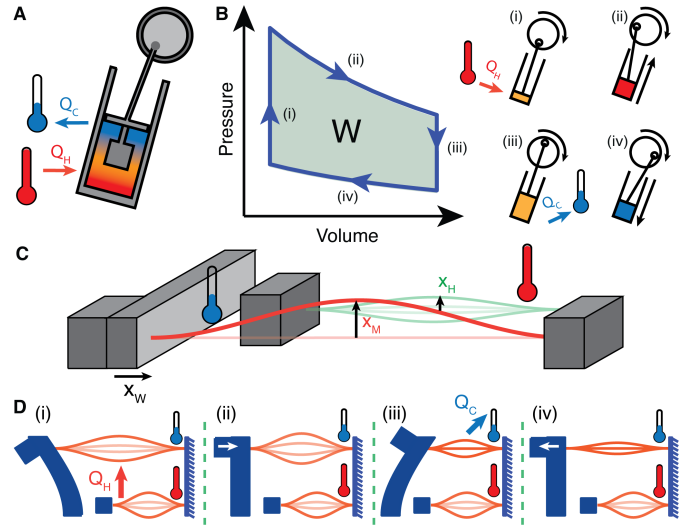

**Figure 1: Cyclic thermal engines**. **A** Stirling heat engine. The engine uses a piston to cyclically compress and expand a gas. A secondary piston displaces the gas and regulates the coupling to the hot and cold reservoirs. **B** Thermal cycle for the Stirling engine. The difference in pressure during expansion and contraction causes the gas to perform net work over a cycle (green shaded area). **C** Mechanical autonomous stochastic engine presented here, consisting of two ribbons, main and secondary (with displacements $x_M$ and $x_H$ respectively) and a cantilever (with displacement $x_W$). **D** Thermal cycle for the proposed engine. This cycle consists of 4 steps: (i) $x_w$ is at its leftmost position and energy flows from $x_H$ to $x_M$ (ii) $M_w$ moves to the right ($\dot{x}_w > 0$), while $x_M$ stays in a high energy state (iii) $x_w$ is at its rightmost position, and energy flows from $x_M$ to the cold bath (iv) $x_w$ moves back to the initial position while $x_M$ stays in a low energy state.

Stirling engines require a mechanism to heat and cool the working fluid in synchrony with the motion of the piston. In conventional engines this can be accomplished by using a secondary internal piston (Fig. 1A) that displaces the fluid and places it in contact with the hot and cold reservoirs. Prior implementations of the stochastic Stirling cycle have used a laser to heat the working particle at pre-determined time intervals[6], making the engine nonautonomous. Our engine attains autonomous operation by utilizing the different resonance responses of the two ribbons. Due to geometric nonlinearity, the resonance frequency of the main ribbon ($f_M$) depends on the position $x_W$ of the cantilever (Fig. 2A), while the resonance frequency of the secondary, hot ribbon ($f_H$) is fixed. As a consequence, the overlap between the respective frequency responses (and therefore the energy transfer[23] between $T_H$ and $x_M$), is controlled by the cantilever (Fig. 2B). By setting the frequency of the main ribbon below the frequency of the secondary ribbon, the maximum energy transfer between the hot thermal bath and the main ribbon occurs when the cantilever is at its rightmost position, as required by the thermal cycle (Fig. 2B). Very recent theoretical proposals in the field of quantum optomechanics utilize a similar mechanism to control the coupling between an optical resonator and a heat source[24,25].

This tension-mediated feedback mechanism introduces the synchronous heating and cooling required for thermal engine operation without the need to externally prescribe periodic temperature variations as in prior works[4,6,7]. The resulting changes in the main ribbon's vibrational energy can be seen in the probability distribution of its position, which is modulated by the cantilever motion (Fig. 2C). The modulation is maximal when the natural frequencies are chosen such that modulation sidebands of the main ribbon's motion coincide with a resonance peak of the coupled system. For weakly coupled ribbons, this condition is approximately $f_H - f_M \approx f_W$, where $f_W$ is the resonance frequency of the cantilever.

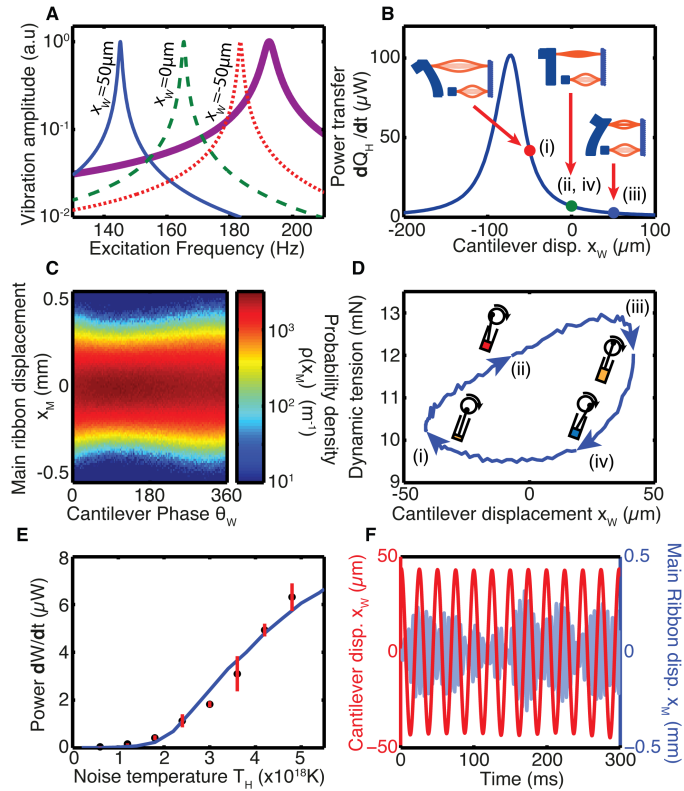

**Figure 2**: **Thermal engine operation**. **A** Uncoupled frequency response of the main ribbon ($x_M$), for cantilever displacements $x_W$ of $-50$ $\mu m$ (blue, solid), $0$ $\mu m$ (green, dashed) and $50$ $\mu m$ (red, dotted). The uncoupled frequency response of the secondary ribbon $x_H$ (thick purple) is shown for comparison. **B** Energy transfer $Q_H$ between the hot bath (applied to the secondary ribbon $x_H$) and the main ribbon ($x_M$), as a function of the cantilever displacement. The colored dots correspond to the curves in **A**. The roman numerals indicate the step of the thermal cycle associated to each displacement and energy transfer. **C** Probability distribution of $x_M$ as a function of the cantilever's oscillation phase, $\theta(\dot{x}_W, x_W)$. **D** Force acting on the cantilever, as a function of the cantilever displacement. **E** Theoretical (blue line) and experimental (black dots) power transfer from the main ribbon to the cantilever as a function of the effective temperature of $x_H$. **F** Time evolution of the cantilever (dark red) and ribbon (light blue).

Figure 2D shows the dynamic tension exerted by the main ribbon on the cantilever as a function of the cantilever's position. This figure has been calculated from the measured probability distribution of the ribbon's position $x_M$ using the equation $T = \gamma <x_M^2>$, and is analogous to the pressure-volume relation in a conventional piston engine. The area inside the curve corresponds to the average work transferred to the cantilever per cycle of operation. The calculated area has a value of $179.2 \pm 7.7$ $nJ$, and it is in good agreement with the average power dissipated in the cantilever, determined to be $171.2 \pm 7.1$ $nJ$ per cycle from the quality factor and average vibrational amplitude of the cantilever (see Supplementary Information). In the experimental range, the power output increases nonlinearly with applied noise temperature (Fig. 2E), and its normalized value $P = 0.095 \pm 0.009$ $k_b T_H s^{-1}$ is comparable to that of stochastic engines reported in the literature, whose values are around $0.02$ $k_b T s^{-1}$[6] and $5$ $k_b T s^{-1}$ [4]. The high effective noise temperatures in Fig. 2E are a consequence of the macroscopic dimensions of our tabletop setup, which mandate the use of an external noise excitation. In the Supplementary Information, we present simulations on a microscopic engine that is capable of producing a measurable work output at a temperature ~200ºC with a predicted output power above $12000$ $k_b T s^{-1}$, owing to its high frequency of operation.

During thermal machine operation, the trajectory of the cantilever is approximately a harmonic signal with a slowly varying envelope, while the motion of the ribbon is highly random (Fig. 2F). We further investigate the properties of the cantilever motion by calculating the phase space probability distribution from the experimental measurements (Fig. 3A) and theoretical simulations (Fig. 3B). This distribution is similar to the theoretically predicted Wigner function for a quantum optomechanical heat engine[25], and contrasts with the Gaussian distribution characteristic of harmonic oscillators subject to a white noise excitation [26]. We compare the phase space distribution during thermal machine operation (Figs. 3A and 3B) with a detuned system by setting $f_H < f_M$ (see Figs. 3C and 3D and Supplementary Information). In the detuned system the cantilever phase space probability follows a Gaussian distribution, which is the distribution that maximizes the entropy for a given mean energy.

We quantify the randomness of the cantilever's motion by calculating entropy of the associated phase space probability distribution (Fig. 3E). The difference between the cantilever's entropy and the corresponding equilibrium entropy increases at high cantilever vibrational energies. This indicates the coexistence of two energy transfer mechanisms: An incoherent mechanism analogous to heat transfer[27], where fluctuations of the main ribbon introduce fluctuations on the ribbon's tension that cause the cantilever to move randomly, and a coherent mechanism where the motion of the main ribbon is modulated by the vibration of the cantilever. At low amplitudes (Fig. 3E and Supplementary Animation 1), or when the main ribbon frequencies are not tuned to result in thermal machine operation (Figs. 3C and 3D), the incoherent mechanism dominates, resulting in a maximally entropic (passive) probability distribution for the cantilever. At high amplitudes the coherent mechanism becomes significant and the motion of the cantilever is nonpassive, with entropy below the maximal value (Fig. 3E). The Fourier transform of the cantilever velocity (Fig. 3F) reveals that the motion occurs mostly at its first mode of resonance. Frequency components corresponding to higher cantilever modes of vibration and to the resonances of the ribbons are below this fundamental component by at least 80 dB. We attribute the presence of small quantities of harmonics to nonlinearities in our measurement system.

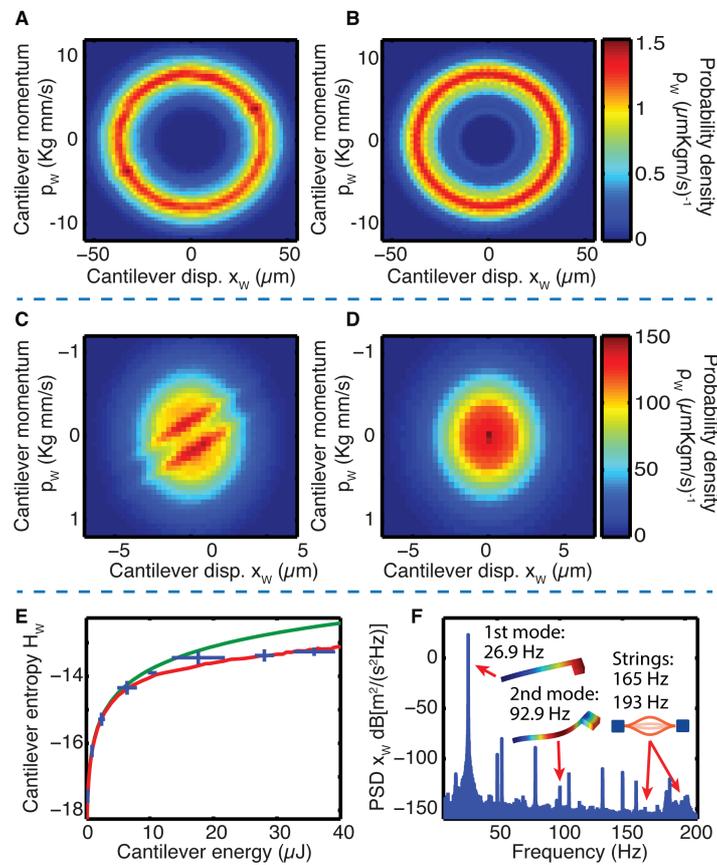

**Figure 3**: **Properties of the cantilever motion**. **A** Experimental phase space probability distribution corresponding to the case when the frequency of the ribbons is tuned to achieve thermal engine operation. **B** Theoretical phase space probability distribution for the experimental case in **A**. **C** Phase space probability density function for the cantilever in the detuned system (where $f_M > f_H$). **D** Theoretical prediction for the system in **C**. **E** Entropy of the cantilever motion as a function of the energy (blue crosses), compared to a theoretical prediction (red line) and to the maximal entropy for the given energy (green line). **F** Fourier transform of the cantilever motion.

The heat engine presented in this Report corresponds to the mass-spring model in Fig. 4A and is described by the system of underdamped [28] Langevin [29,30] equations (See supplementary information for derivation and numerical algorithms):

$$m_H\ddot{x}_H + b_H\dot{x}_H + k_Hx_H + k_{HM}(x_H - x_M) + \mu x_H^3 = \xi_H$$
$$m_M\ddot{x}_M + b_M\dot{x}_M + (k_M - 2\gamma x_W)x_M + k_{HM}(x_M - x_H) + \mu x_M^3 = \xi_C$$
$$m_W\ddot{x}_W + b_W\dot{x}_W + k_Wx_W - \gamma x_M^2 = \xi_W$$

Eq. 1 A-C

Here $k_H$, $k_M$ and $k_W$ are the stiffness of the hot ribbon, the main ribbon and the cantilever respectively, $b_H$, $b_M$ and $b_W$ correspond to the damping acting on these respective degrees of freedom and $m_H$, $m_M$, and $m_W$ are the corresponding masses. The terms $\xi_H$, $\xi_C$ and $\xi_W$ represent the thermomechanical Johnson-Nyquist noise introduced by the thermal baths acting on each degree of freedom. These terms have a white noise power spectral density of $4K_BT_Xb_X$. The constants $k_{HM}$ represent the linear coupling between the two ribbons, and $\gamma$ the nonlinear coupling between the main ribbon and the cantilever. Coupled degrees of freedom subject to Brownian motion have been studied in electronic systems[27], and the asymmetric coupling between $x_M$ and $x_W$ appears in the description of phonon modes in superconductors[31].

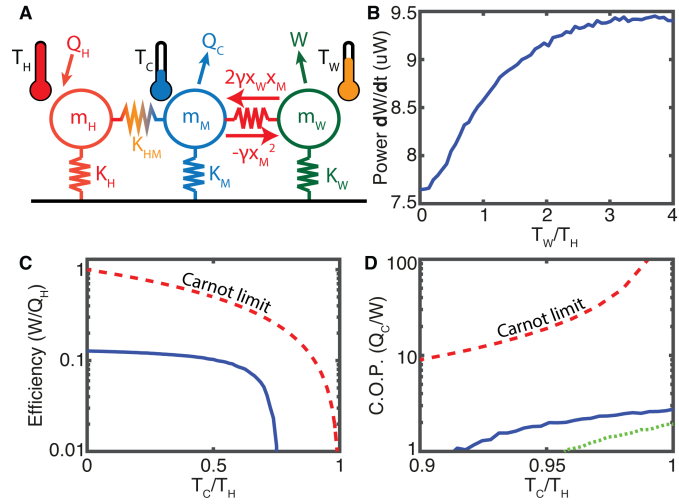

**Figure 4**: **Theoretical investigation**. **A** Mass-spring model for the system. **B** Energy transfer as a function of the cantilever temperature $T_W$. **C** Efficiency of the thermal machine (blue, solid) and comparison with the Carnot efficiency (red, dashed). Here, the cantilever motion has been externally prescribed to be $x_W = (50\ \mu m)\cos(\omega_W t)$ in order to prevent incoherent energy transfer between the ribbons and the cantilever. **D** Refrigerator Coefficient Of Performance ($C.O.P. = Q_C/W$) when the cantilever is forced to oscillate at $A_W = 50\ \mu m$ (blue, solid) and when driven by noise at $T_W = k_W A_W^2/k_B$ (green, dotted). The red dashed line is the Carnot maximal C.O.P (red).

As demonstrated in Fig. 3, the theoretical model predictions are in good agreement with the experiments. Thus, we use the model to determine quantities that are not directly measurable in our experimental setup, such as the energy transfer between ribbons. We highlight the most relevant theoretical findings in Fig. 4. When the temperature of the bath surrounding $x_W$ is increased the energy transferred between $x_H$ and $x_W$ increases (Fig 4A). This corresponds to an effective negative thermal conductivity. In addition, the system is still able to transfer energy between the thermal baths $T_H$ and $T_W$ when $T_W$ is increased above $T_H$. This observation, which seems to defy the second law of thermodynamics, is made possible by the fact that $x_M$ is at a lower temperature than $x_H$, and absorbs the excess entropy extracted from $\xi_H$. Figure 4C presents the efficiency of the thermal machine as a function of the ratio between $T_H$ and $T_C$. The machine attains a maximum efficiency of approximately 30% of the maximal Carnot efficiency using our experimental, unoptimized parameters. In the Supplementary Information, we present alternative designs that attain efficiencies up to 50% of the theoretical maximum. When the temperature ratio $T_H/T_C$ is close to 1, the energy flow between $T_H$ and $T_C$ reverses, and the machine acts as a refrigerator [17]. The refrigerator regime requires a constant supply of energy to the cantilever. This energy can be provided by externally prescribing the cantilever displacement, or by increasing $T_W$ to introduce large amplitude thermal motion in the cantilever. In the latter case, the main ribbon is effectively being cooled through the addition of heat to the system, which behaves as an absorption refrigerator [32-35]. Figure 4D presents the efficiency of the refrigerator operation.

This work has demonstrated that a mechanical system consisting of two ribbons and a cantilever has the ability to act as a heat engine or refrigerator, and presents the unusual property of negative thermal conductivity. Traditionally, Brownian motion has been seen has an inconvenience when present in mechanical systems, e.g., by limiting the precision of nanomechanical sensors[36]. Our work demonstrates that this thermomechanical noise is a tool to study thermodynamics in both macro and micro scale systems, and has the potential to be used as a source of energy.

## Acknowledgments

This work was supported by ETH Research Grant ETH-24 15-2.

# Supplementary Information: A mechanical autonomous stochastic heat engine

Marc Serra-Garcia*[1], André Foehr[1], Miguel Moleron[1], Joseph Lydon[1], Christopher Chong[2] and Chiara Daraio[1,3]

[1]Department of Mechanical and Process Engineering, Swiss Federal Institute of Technology (ETH), Zürich, Switzerland

[2]Department of Mathematics, Bowdoin College, Brunswick, ME 04011, USA

[3]Engineering and Applied Science, California Institute of Technology, Pasadena, California 91125, USA

*email: sermarc@ethz.ch

# Alternative heat engines

In the main paper it was demonstrated that a mechanical system can act as a stochastic heat engine by exploiting nonlinearity and resonance. The generality of this fundamental concept is explored in this section. We briefly discuss two alternative arrangements of masses and nonlinear springs that also function as heat engines.

## Main-Cold engine

We first consider an engine similar to the engine that was discussed in the main paper (Fig. S1A). The only difference is that the hot thermal bath is applied to the main ribbon instead of the secondary ribbon. For the engine to operate, the resonance frequency of the main ribbon must be above the natural frequency of the secondary, cold ribbon (Fig. S1B).

The engine is described by the system of equations:

$$m_C \ddot{x}_C + b_C \dot{x}_C + k_C x_C + k_{CM}(x_C - x_M) + \mu x_C^3 = \xi_C$$
$$m_M \ddot{x}_M + b_M \dot{x}_M + (k_M - 2\gamma x_W) x_M + k_{CM}(x_M - x_C) + \mu x_M^3 = \xi_H$$
$$m_W \ddot{x}_W + b_W \dot{x}_W + k_W x_W - \gamma x_M^2 = \xi_W$$

Eq. S1 A-C

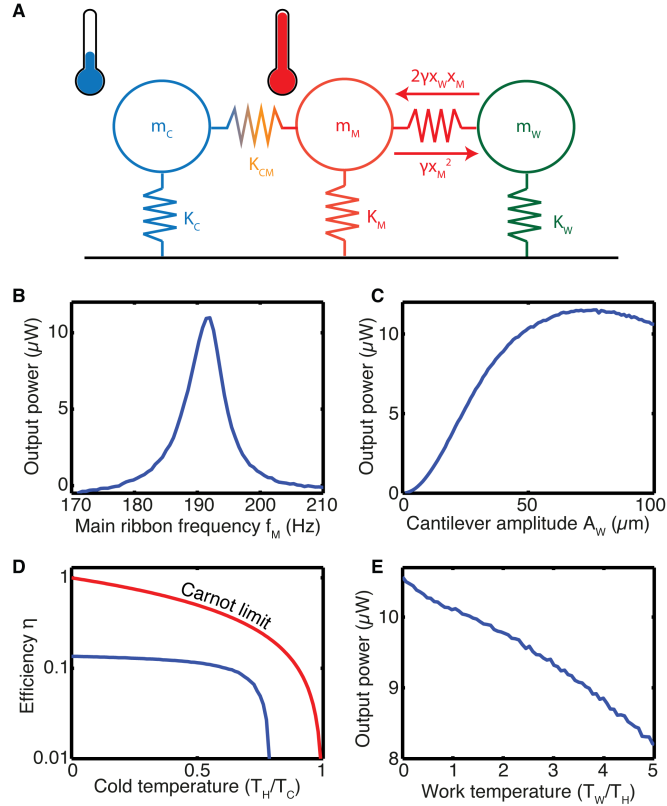

**Supplementary Figure 1: Main-Cold engine**: **A** Mass-spring diagram of the engine. **B** Output power as a function of the main ribbon's frequency, for a prescribed cantilever vibration amplitude of 50 $\mu m$. **C** Engine power as a function of the work amplitude ($dW/dt = < -\gamma x_M^2 \dot{x}_W >$). **D** Energy efficiency and comparison with the Carnot limit. **D** Power output as a function of the temperature acting on the cantilever. The parameters used in this simulation: $f_M = 190.8\ Hz$, $f_C = 165.37\ Hz$, $f_W = 26.87\ Hz$, $m_M = m_C =$

$0.207 g$, $m_W = 1.27\ Kg$, $Q_C = 59.41$, $Q_M = 167.78$, $k_{CM} = 0.0381 \cdot k_C$, $\gamma = 513\ kNm^{-2}$, $\mu = 16.9\ MNm^{-3}$, $T_H = 2 \cdot 10^{18} K$, $T_M = T_C = 0$, $A_W = 50\ \mu m$ (unless otherwise indicated).

The efficiency and output power of this engine are similar to those obtained in the main paper (Fig. S1 C-D), where the hot bath was applied to the secondary ribbon. This engine is also capable of pumping energy from a cold thermal bath to a hot thermal bath (Fig. S1 E). However, in contrast with the engine presented in the main paper, in this case an increase in the cantilever temperature decreases the power transferred to $x_W$.

## Hot-Main-Cold engine

We can build engines with higher efficiency and output power by adding an additional ribbon (Fig. S2A). In this case, the main ribbon ($x_M$) is attached to two ribbons. One of them ($x_H$) is in contact with the hot thermal bath, while the other ($x_C$) is in contact with the cold thermal bath. For maximum efficiency, the hot ribbon's natural frequency should be tuned above the frequency of the main ribbon, while the cold ribbon's natural frequency has to be set below the frequency of the main ribbon (Fig. S2B). This system is described by the system of equations:

$$m_H \ddot{x}_H + b_H \dot{x}_H + k_H x_H + k_{HM}(x_H - x_M) + \mu x_H^3 = \xi_H$$
$$m_C \ddot{x}_C + b_C \dot{x}_C + k_C x_C + k_{CM}(x_C - x_M) + \mu x_C^3 = \xi_C$$
$$m_M \ddot{x}_M + b_M \dot{x}_M + (k_M - 2\gamma x_W) x_M + k_{CM}(x_M - x_C) + k_{HM}(x_M - x_H) + \mu x_M^3 = \xi_M$$
$$m_W \ddot{x}_W + b_W \dot{x}_W + k_W x_W - \gamma x_M^2 = \xi_W$$

Eq. S2 A-D

Figure S2C shows the power output ($\dot{W} = <F\dot{x}> = <-\gamma x_M^2 \dot{x}_W>$) as a function of the prescribed cantilever vibration amplitude. The relation between cantilever amplitude and output power is similar to that obtained for the Cold-Main engine of the previous section and the Hot-Main engine of the main paper, where a nonlinear increase in output power is observed. In this case, however, the saturation occurs at higher amplitudes, and the output power is larger by a factor of more than two (Compare Figs S1C and S2C).

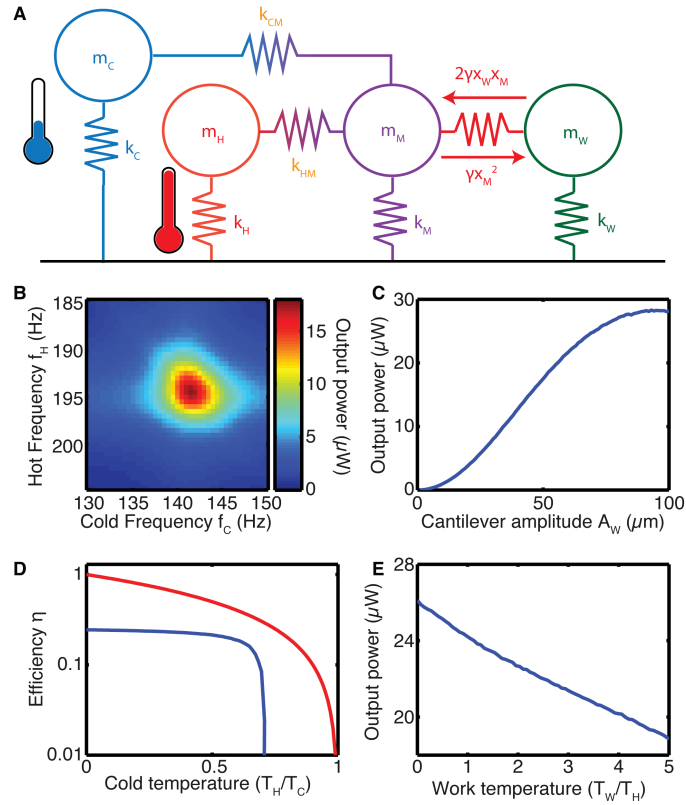

**Supplementary Figure 2: Hot-Main-Cold engine**: **A** Mass-spring diagram of the engine. **B** Engine efficiency as a function of the cold spring stiffness. **C** Engine power as a function of the work amplitude. **D** Energy transfer as a function of the work temperature. **E** Efficiency in the nonautonomous limit. The parameters used in this simulation are: $f_H = 194.5\ Hz$, $f_M = 165.37\ Hz$, $f_C = 141.5\ Hz$, $f_W = 26.87\ Hz$, $m_H = m_M = m_C = 0.207g$, $m_W = 1.27\ Kg$, $Q_H = Q_C = 59.41$, $Q_M = 1000$, $k_{CM} = k_{HM} = 0.0381 \cdot k_C$, $\gamma = 513\ kNm^{-2}$, $\mu = 16.9\ MNm^{-3}$, $T_H = 2 \cdot 10^{18} K$, $T_M = T_C = 0$, $A_W = 50\ \mu m$ (unless otherwise indicated).

In addition to an increase in power, this design also brings an increase in efficiency (Figs. S2 C and D). This efficiency reaches a maximum of 47% of the Carnot limit, which is significantly higher than the efficiency of the two-ribbon engines (about 30%). As in the previous two cases, the engine is able to transfer energy from cold to hot temperatures (Fig S2E).

# Equations of motion for the ribbon-cantilever system

Our system consists of two pieces of metal ribbon ($x_M$ and $x_H$) and a cantilever ($x_W$). In this section we derive the coupled equations of motion for the system (Eq. 1 of the main paper) as a function of the geometry.

## Cantilever equation

The motion of the cantilever can be described by considering only the first bending mode of vibration. This is a consequence of the fact that higher modes of vibration are much stiffer and therefore vibrate at smaller amplitudes. This assumption is confirmed experimentally by inspecting the Fourier transform of the cantilever motion (Fig 3F of the main paper). Under this assumption, the dynamics of the cantilever are described by a damped-driven harmonic oscillator:

$$m_W \ddot{x}_W + b_W \dot{x}_W + k_W x_W = T(x_M) \qquad \text{Eq. S3}$$

where $m_W$, $b_W$ and $k_W$ are the effective mass, damping and stiffness, respectively, of the normal mode with position $x_W$. The right hand side sums the external forces acting on the mode. In our experimental system, the only external force acting on the cantilever is the ribbon tension, which we represent by $T$. The value of $T$ is a function of the ribbon's in-plane vibrational displacement $x_M$ (We will obtain an expression for $T(x_M)$ in the *Ribbon equation* section). The variable $x_W$ represents the position of the normal mode at a particular instant of time. As such, it can be defined in different ways depending on the mode's normalization. In this derivation, we set the mode normalization that makes the modal coordinate $x_W$ coincide with the displacement of the cantilever at the point where the ribbon is attached. We calculate the effective mass and stiffness using the commercial Finite Element Method (FEM) package Comsol Multiphysics®. The motion of the cantilever must take into account the longitudinal stiffness of the ribbon. We use a value of 31.55 $kN/m$ that correctly reproduces the measured cantilever frequency, as described in the section *Experimental determination of the quality factors, frequencies and coupling constant,* where we will also determine the damping constant $b_W$.

The effective mass is defined as the integral $\int_V \rho(x, y, z)(u^2 + v^2 + w^2)dV$ where $u$, $v$ and $w$ represent the displacements of the mode along the directions $x$, $y$ and $z$ respectively, while $\rho$ is the density of the cantilever material. This provides us with a cantilever mass of $1.27 \pm 0.04\ kg$, where the uncertainty originates from tolerances in the geometry measurement. The effective stiffness can be calculated from the eigenfrequency using the equation $k = m\omega^2$.

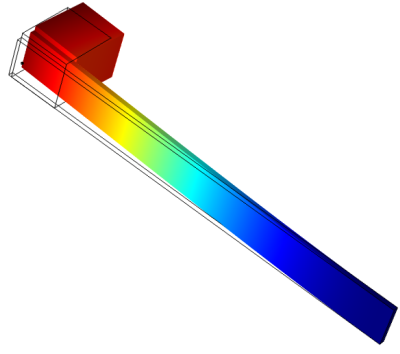

**Supplementary Figure 3: Cantilever's first normal mode of vibration.** We use this mode profile to calculate the effective mass of the cantilever ($m_w$ in our system of equations). The cantilever has a length of 40.6 cm, a height of 4 cm and a thickness of 5 mm. The block mass at the cantilever's end has dimensions of 5x5x6 cm. The cantilever is made of steel with a Young's modulus $E = 200\ GPa$, a Poisson ratio $\nu = 0.33$ and a density $\rho = 7850\ Kg/m^3$.

## Ribbon equation

The motion for a beam-like resonator can be approximated by the Woinowski-Krieger equation[1]. This equation is a modified Euler-Bernoulli equation that takes into account the effect of the vibration-induced tension. The Woinowski-Krieger equation for a slender beam of length $L$ takes the form:

$$EI\frac{\partial^4 w}{\partial y^2} - \frac{EA}{L}\left(\Delta y + \Delta y_0 + \frac{1}{2}\int_0^L \left(\frac{\partial w}{\partial y}\right)^2 dy\right)\frac{\partial^2 w}{\partial y^2} + \rho A\frac{\partial^2 w}{\partial t^2} - f(y,t) = 0 \quad \text{Eq. S4}$$

Here, $\rho$ is the density of the ribbon, $A$ is the cross-section area, $I$ is the bending moment of inertia, and $E$ is the Young's modulus. The $y$ axis corresponds to the direction along the length of the ribbon, and $w$ is the deflection. $\Delta y_0$ is the elongation of the ribbon when the cantilever is at equilibrium. The displacement of the cantilever from its equilibrium position introduces a $\Delta y$ change in the length of the ribbon.

At small amplitudes, the Woinowski-Krieger's nonlinear term (represented by the integral in Eq. S4) vanishes. In this regime, the system behaves like an Euler-Bernoulli beam in which case it would be reasonable to define a set of vibrational modes for the ribbon. It is possible to extend the range of validity of this picture to higher amplitudes by first performing a modal decomposition and then using the Galerkin method to account for the nonlinear effects[2]. This approach is valid as long as the amplitude is small enough to ensure that the mode profile is not significantly affected by the nonlinearity. Here we use this method to derive a nonlinear equation for the motion of the first mode of vibration for the ribbons. We follow the same approach as Postma et al. [2]. This approach consists in expressing the solution of Eq. S4 as a time-dependent linear combination of trial functions:

$$w(y,t) = \sum_i^n u_i(t)\phi_i(y) \quad \text{Eq. S5}$$

In general, a finite set of trial functions will not be able to exactly satisfy Eq. S4. This implies that the left hand side of the equation will not be identically zero. Instead, there will be an error term $\epsilon(x,t)$:

$$EI\frac{\partial^4 w}{\partial y^2} - \frac{EA}{L}\left(\Delta y + \Delta y_0 + \frac{1}{2}\int_0^L \left(\frac{\partial w}{\partial y}\right)^2 dx\right)\frac{\partial^2 w}{\partial y^2} + \rho A \frac{\partial^2 w}{\partial t^2} - f(y,t) = \epsilon(y,t) \qquad \text{Eq. S6}$$

The Galerkin method imposes the condition that the error term should be orthogonal to each of the trial functions. Orthogonality between $\epsilon(x,t)$ and $\phi_i(x)$ is defined as:

$$\int_0^L \phi_i(t)\epsilon(y,t)\,dy = 0 \qquad \text{Eq. S7}$$

The system is completely determined because there are as many orthogonality conditions as unknown functions $u_i(t)$. We use the eigenmodes of the linear system as test functions $\phi_i(y)$, and we consider only the first normal mode of vibration. This assumption is reasonable since higher order modes are considerably stiffer, and therefore oscillate at much smaller amplitudes. Applying the orthogonality condition (Eq. S7) to Eq. S6 results in the following ordinary differential equation describing the time evolution of the mode coordinate $u_1$ (Henceforth referred to as $u$):

$$m\ddot{u} + ku + 2\beta\Delta u + \mu u^3 = F(t) \qquad \text{Eq. S8}$$

The term $2\beta\Delta u$ is responsible for the coupling between the cantilever and the ribbon. The parameters in Eq. S8 are given by:

$$m = \rho A \int_0^L \phi^2(y)dy \qquad k = EI\int_0^L \frac{\partial^4 \phi(y)}{\partial y^4}\phi(y)dy + 2\beta\Delta y_0$$

$$2\beta = -\frac{EA}{L}\int_0^L \frac{\partial^2 \phi(y)}{\partial y^2}\phi(y)dy \qquad \mu = \beta \int_0^L \left(\frac{\partial \phi(y)}{\partial y}\right)^2 dy$$

Eq. S9 A-D

In order to evaluate these expressions, we calculate the mode profile $\phi(y)$ using the pre-stressed eigenfrequency analysis function of Comsol Multiphysics® 5.0. We use the plate model to describe the dynamics of the ribbon. In the finite element simulation, we select an elongation $\Delta x_0 = 250\mu m$ that yields resonance frequencies that are comparable to the experimental values. The thickness of the beams is set to $16.25 \pm 1.83\ \mu m$ in order to reproduce the longitudinal stiffness of $31.55 \pm 1.04\ kN/m$ that yields the correct cantilever resonance frequency, assuming a density $\rho = 8500\ kg/m^3$ and a Young's modulus $E = 115 \pm 10\ GPa$. The integration in Eq. S9a-d is performed along the central axis of the beam. For convenience, we fix the normalization of the mode to be:

$$\int_0^L \phi^2(y)dy = L \qquad \text{Eq. S10}$$

This definition sets the effective mass $m = \rho A L$ equal to the physical mass of the ribbon. The values of the parameters corresponding to our experiment are given in Table S1:

| Parameter | Value |
| --- | --- |
| m | $0.207 \pm 0.023\ g$ |
| $\beta$ | $513 \pm 56\ kN/m^2$ |
| $\mu$ | $16.9 \pm 1.8\ MN/m^3$ |

**Supplementary Table 1:** Cantilever parameters.

We will calculate the value of $\Delta y_0$ and $k$ from the frequency response in the section *Experimental determination of the quality factors, frequencies and coupling constant.*

### Coupling between ribbons

The coupling between $x_H$ and $x_M$ is implemented by a small piece of ribbon placed between the two resonators. The coupling is assumed to be linear, and modeled with a term $F = k_{HM}(x_H - x_M)$. We will experimentally measure the value of $k_{HM}$ in the following section.

## Experimental determination of the quality factors, frequencies and coupling constant

The previous section describes a model for the system that is based on analytical and finite element simulations. These methods are adequate to predict several of the system parameters. In this section we experimentally measure the remaining system parameters that cannot be accurately predicted with theoretical models.

### Cantilever frequency and quality factor

We measure the cantilever by exciting it using a harmonic signal and measuring the frequency response. The excitation is provided by a solenoid that induces a force on a small magnet attached to the cantilever. The frequency response is extracted using a Lock-In amplifier that receives the signal from a Laser Doppler Vibrometer. Supplementary Figure 4 shows the measured frequency response and the corresponding fit using the MATLAB® state space parameter estimation function. The fit yields a cantilever resonance at $26.87 \pm 0.01$ Hz with a quality factor of $964 \pm 41$.

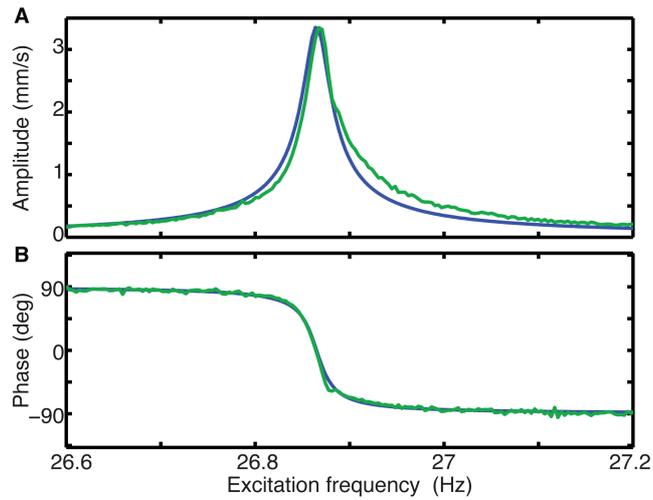

**Supplementary Figure 4: Cantilever's frequency response. A** Amplitude of the cantilever's velocity as a function of the excitation frequency. **B** Phase of the cantilever's velocity as a function of the excitation frequency. The green lines represent measured values, while the blue lines represent the fitted response.

## Ribbon's resonance frequencies, couplings and damping

In this section, we determine the experimental quality factor, frequency and coupling stiffness for the ribbons. We accomplish this by measuring the frequency response of the ribbon when excited by a harmonic signal of variable frequency. The results are fitted using MATLAB® state space parameter estimation.

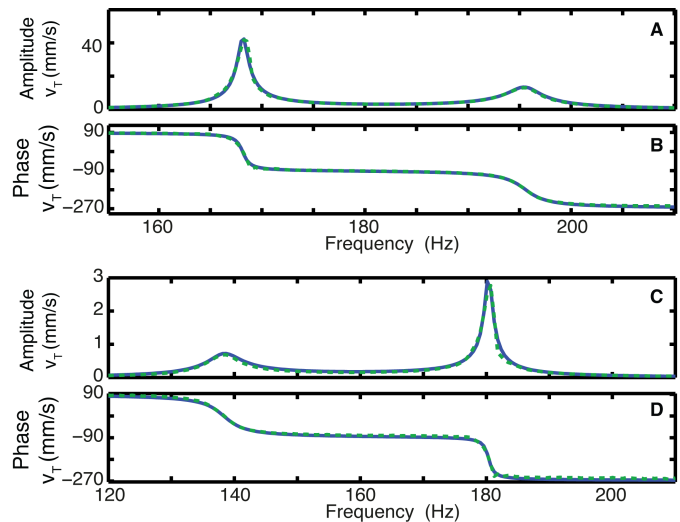

**Supplementary Figure 5: Ribbon frequency response measurement. A, B:** Ribbon amplitude (**A**) and phase (**B**) as a function of the excitation frequency, for the case where the ribbon tensions are tuned to result in thermal machine operation (Fig. 3 A, B, D, E in the main paper). **C,D:** Amplitude (**C**) and phase (**D**) as a function of the excitation amplitude for the case where the ribbon are detuned so thermal machine operation does not occur (Fig. 3 C, D in the main paper). In all four panels, the dashed green line represents the experimental data and the solid blue line represents the fitted response.

The state space parameter estimation results are given in Table S2.

| Parameter | Value (Tuned) | Value (Detuned) |
|---|---|---|
| $f_H$ | $192.55 \pm 0.02\ Hz$ | $134.32 \pm 0.04\ Hz$ |
| $f_M$ | $165.37 \pm 0.02\ Hz$ | $176.67 \pm 0.02\ Hz$ |
| $Q_H$ | $59.3 \pm 1.8$ | $26.9 \pm 1.2$ |
| $Q_M$ | $167.8 \pm 3.0$ | $151.7 \pm 2.0$ |
| $k_{HT}/k_M$ | $0.0381 \pm 0.0059$ | $0.0381 \pm 0.0060$ |

**Supplementary Table 2:** Ribbon parameters.

We attribute the large differences between $Q_H$ and $Q_M$ this to eddy current losses in the ribbon due to the large magnetic fields used for excitation.

# Methods

## Numerical methods

We simulate the system by expressing the equations of motion in Itô form:

$$dx_H = v_H \cdot dt + 0 \cdot dW_1$$
$$dx_M = v_M \cdot dt + 0 \cdot dW_2$$
$$dx_W = v_W \cdot dt + 0 \cdot dW_3$$
$$dv_H = -\frac{1}{m_H}\big(b_H v_H + k_H x_H + k_{HM}(x_H - x_M)\big) \cdot dt + \sqrt{2 k_B b_H T_H} \cdot dW_4$$
$$dv_M = -\frac{1}{m_M}\big(b_M v_M + (k_M - 2\gamma x_W) x_M + k_{HM}(x_M - x_H)\big) \cdot dt + \sqrt{2 k_B b_M T_M} \cdot dW_5$$
$$dv_W = -\frac{1}{m_M}\big(b_W v_W + k_W x_W - \gamma x_M^2\big) \cdot dt + \sqrt{2 k_B b_W T_W} \cdot dW_6$$

Eq. S11 A-F

In this system, the terms $m_H, m_M, m_W, k_H, k_M, k_W, k_{HM}, b_H, b_M, b_W$ and $\gamma$ have the same meaning as in Eq. S1 in the main paper. The excitation temperature in the simulations was fitted in order to reproduce the experimental results. This also required adjusting the natural frequency of the secondary ribbon by $+0.5\ Hz$. The vector $dW$ represents the thermomechanically induced velocity change incurred during the time interval $dt$, and has a Gaussian distribution.

The Gaussian-distributed random numbers were obtained using a Mersenne-Twister uniform random number generator and the Beasley-Springer-Moro inversion formula. We simulated the time evolution of the system using a stochastic Runge-Kutta algorithm with strong order 1.5[3] implemented in C++, which ran on the ETH Euler cluster. Statistical quantities such as the RMS velocities were averaged over an interval of 5000 s, using a time step of $3\ \mu s$ for all the figures except the microscopic engine. The microscopic simulations use a time step of $60\ ps$ and an averaging time of $0.1\ s$.

## Experimental materials and methods

We measured the cantilever motion using a Polytec Laser Doppler Vibrometer (LDV), model OFV-503 with a decoder model OFV-3001 placed at an angle of 45° (Fig. S6A). For the ribbon, we used a Polytec LDV model OFV-505 with a decoder model OFV-5000, pointing at $L = 14.1\ cm$ along the length of the ribbon. We calculated the modal amplitude from the measured velocity by using the mode profile function obtained from finite element simulation. The sensitivity of the lasers was set to 125 $mms^{-1}V^{-1}$ for the cantilever and 200 $mms^{-1}V^{-1}$ for the ribbon. We digitized the signal from the lasers for a length of 2000s of 2.5 $ksamples/s$ using a Tektronix™ oscilloscope model DPO-3034 in high-resolution mode. We calculated the displacements of the cantilever and ribbon by low-passing the velocity signal to remove any DC component and then integrating the resulting signal using the MATLAB® cumtrapz function. The cutoff frequencies for the high-pass filter were selected at 5 Hz for the cantilever and 60 Hz for the ribbons.

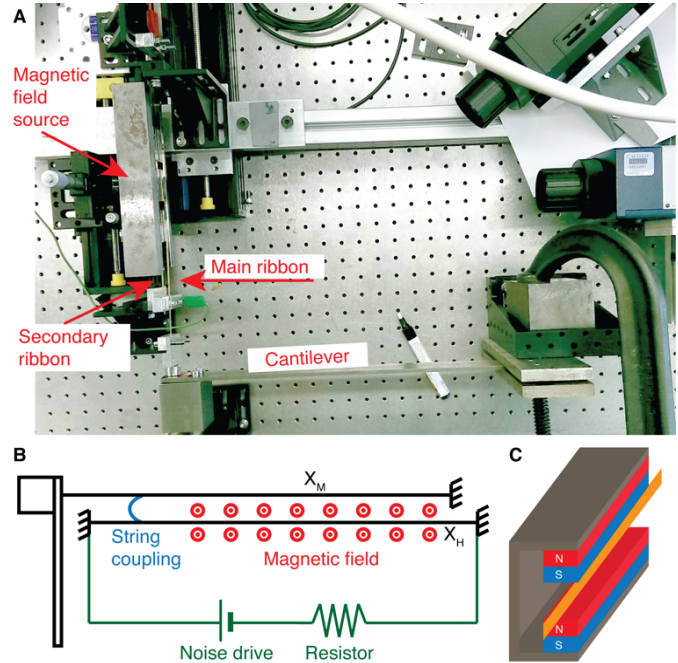

**Supplementary Figure 6: Experimental setup. A** Picture of the experimental setup with the two ribbons and the cantilever. **B** Schematic diagram of the experimental setup including the noise excitation. **C** Steel structure and magnets used to apply a noise excitation to the secondary ribbon.

We implemented the ribbons by cutting a sheet of brass obtained from Brütsch-Rügger™ (Catalog number 162950.0110) with a nominal thickness of $20 \pm 2\ \mu m$ into two stripes of height $H = 5\ mm$ and length $L = 50\ cm$. The free length the ribbons (after clamping) was $L_R = 30\ cm$. We placed the two ribbons at a distance of 8 $mm$ and introduced a coupling between them by soldering an additional strip of brass at both ribbons at length $L_C = 6\ cm$ along their length (Fig. S6B).

We introduced the noise excitation on the secondary ribbon by placing it under a magnetic field implemented with 12 neodymium magnets from Supermagnete™ catalog number Q-40-10-05-N enclosed in a C-shaped steel bar (Fig. S6C). We circulated the noise current across the ribbon by driving both ends from a Topping TP22 amplifier in series with a 5.5 Ohm power resistor. The noise signal driving the ribbon was obtained by generating a sequence of normally distributed random numbers, which was subsequently lowpassed using an 8th order Low-Pass filter with a cutoff frequency of 10 KHz. The electrical signal was generated with the sound card of a MacBook laptop using Apple™ CoreAudio library. The Fourier transform of the amplified signal revealed that the noise spectrum was flat on the region of interest between 50 Hz and 300 Hz.

*Determination of statistical quantities and errors*

We calculated the histograms in Figure 2D and Figure 3 A-D by binning the time series data from the numerical simulations into a two-dimensional histogram with 128x128 bins. We calculated the entropies from the histograms using the definition:

$$H = \int_\Gamma \rho(x_W, p_W) \log \rho(x_W, p_W)\, dx_W dp_W \qquad \text{Eq. S12}$$

Where $\Gamma$ is the phase space of the cantilever, $x_W$ is the cantilever position and $p_W$ is the cantilever momentum. The statistical uncertainties associated to each time series measurement were calculated by dividing the measured time series in 8 sub-sequences of equal length. We then evaluated the variance of the measured quantities in the 8 sub-sequences and utilized this value to extrapolate the uncertainty of our measurement.

# Microscopic heat engine operating on thermal fluctuations

The experimental thermal engine that we have presented in this paper requires an external noise source to simulate a large temperature and therefore obtain significant amounts of Brownian motion. This is a consequence of the engine's macroscopic size, since Brownian motion in a meter-sized steel structure is extremely small. The goal of this section is to demonstrate that it is possible to implement engines producing measurable amounts of work at experimentally accessible temperatures (~200ºC) by miniaturizing the ribbons and the cantilever. The dimensions of our microscopic engine are experimentally feasible, and we base our simulations in parameters obtained from the literature, corresponding to actual experimental devices.

We model the cantilever using a diamond (Young's modulus $E_C = 960\ GPa$ and density $\rho_C = 3500\ Kg/m^3$) beam of length $L_C = 20\ \mu m$, width $w_C = 880\ nm$ and height $h_C = 370\ nm$, with a quality factor $Q_C = 47800$ (See [4] for details on the geometry, quality factor and fabrication process). The ribbons are made with graphene and have a length $L_R = 2.7\ \mu m$, width $w_R = 20\ nm$, thickness $t_R = 3\ nm$ and quality factor $Q_R = 35$. Graphene ribbons with this geometry can be modeled using conventional finite element theory or beam theory (See Fig. 2D in Ref. [5]), using a Young's modulus of $E_R = 1\ TPa$ and a density $\rho_R = 2200\ kg/m^3$.

The main ribbon is attached to the cantilever at a distance $x_a = 2.9\ \mu m$ from the cantilever's supporting end (Fig. S7A). This is because the longitudinal stiffness of the ribbon is very high and would place an excessive load to the cantilever if it were attached at the end. The equations of motion for the microscopic system are:

$$m_H \ddot{x}_H + b_H \dot{x}_H + k_H x_H + k_{HT}(x_H - x_T) + \mu x_H^3 = \xi_H \qquad \text{Eq. S13}$$
$$m_M \ddot{x}_M + b_M \dot{x}_M + (k_M - 2\gamma_0 \phi(x_A) x_W) x_M + k_{HM}(x_M - x_W) + \mu x_C^3 = \xi_C \qquad \text{A-C}$$

$$m_W \ddot{x}_W + b_W \dot{x}_W + k_W x_W - \gamma_0 \phi(x_A) x_M^2 = \xi_W$$

Where $x_W$ represents the displacement of the cantilever at $x = L_C$, $\phi(x_a)$ is the cantilever's first bending mode profile, evaluated at the attachment point of the ribbon. The ribbon's nonlinear parameter $\gamma_0$ can be computed using Eq. S9, and all the other symbols have the same meaning as in Eq. 1 of the main text. Equation S13 can in fact be reduced to Eq. 1 by defining $\gamma = \gamma_0 \phi(x_A)$. This definition provides us with a mechanism to adjust the value of $\gamma$ when designing a device.

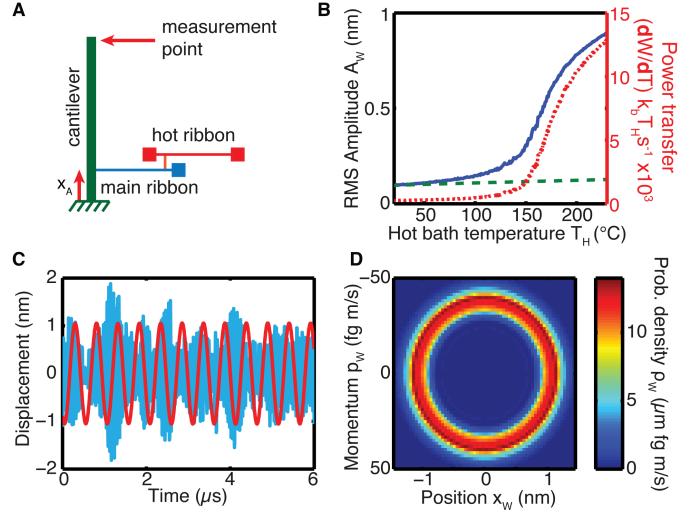

**Supplementary Figure 7: Microscopic engine. A** Schematic diagram of the engine, with the hot ribbon attached at a distance $x_A$ from the cantilever's support. **B** Cantilever's vibration amplitude as a function of the hot thermal bath temperature $T_H$ (blue, solid) compared to the thermomechanical amplitude of the cantilever at $T_H$ (green, dahsed). The red dotted line is the output power, which reaches a maximum value of 12810 $k_B T_H s^{-1}$ at 230 °C. **C** Example of a cantilever's trajectory (red) and a graphene nanoribbon's trajectory (light blue). Phase space probability distribution for the cantilever, showing a distinctively nonpassive circular shape.

Figure S7B presents a simulation of the thermal machine's cantilever amplitude as a function of the hot thermal bath temperature. In all simulations, both the cold bath temperature $T_C$ and the cantilever's temperature $T_W$ are set to 20 °C. The RMS value of the cantilever's displacement $x_W$ is much larger than it can be attributed to thermomechanical noise. If the cantilever's temperature were to raise to $T_H = 180°C$ due to direct conduction from the cold bath, the thermomechanical noise would still be more than 6 times smaller than the simulated vibration amplitude. At a temperature $T_W = 180°C$, the cantilever reaches a peak amplitude of 1 $nm$ (Fig 7C). This motion can be measured experimentally at very high resolution (See Fig. 3 in Ref. [4]). Therefore, the thermal machine effect should be observable in a system of these dimensions. Due to its high operating frequency, the proposed engine is capable of producing an output power orders of magnitude above that of prior works.

We have shown that a microscopic autonomous thermal engine can produce a detectable amount of power. Our calculations use realistic parameters obtained from the literature. However, constructing a device with these properties is challenging and our description omits some technical aspects: Precisely controlling the attachment point of a graphene ribbon is a hard nanofabrication problem. In addition, the device requires a mechanism to heat the hot ribbon, and to tune the frequency response. The heating can be accomplished by passing a current through the ribbon, or by utilizing thermally conductive electrodes. The frequency tuning may be accomplished by imposing a static deflection on the cantilever by means of an electrostatic potential. Finally, the required quality factors are high. This latter aspect can be mitigated by numerically optimizing the design (Or switching to a more powerful alternate engine such as the one discussed in Fig. S2). Increasing the temperature while decreasing the operating amplitude and stiffness of the cantilever will also result in a relaxed quality factor requirement.

### System parameters

The parameters used for the numerical simulations of the microscopic engine are given in Table S3.

| Parameter | Value |
|---|---|
| $f_M$ | $27.203\ MHz$ |
| $f_M$ | $30.195\ MHz$ |
| $\beta$ | $4.47 \cdot 10^7\ N/m^2$ |
| $\mu$ | $1.8 \cdot 10^{14}\ N/m^3$ |
| $m_H = m_M$ | $3.56 \cdot 10^{-19}\ Kg$ |
| $m_W$ | $2.85 \cdot 10^{-15}\ Kg$ |
| $k_{HM}$ | $0.0938\ k_M$ |
| $Q_W$ | $47800$ |
| $Q_H = Q_M$ | $35$ |

**Supplementary Table 3:** Microscopic engine parameters.

These parameters have been obtained using the same procedure that we described in the section *Equations of motion for the ribbon-cantilever system*. The FEM simulations on the cantilever account for the stiffness of the ribbon by using a spring foundation at the ribbon attachment point. The ribbon's frequencies require a tension of 2.2 $nN$ for the main ribbon and 2.9 $nN$ for the hot ribbon.